\documentclass[10pt]{article}
\usepackage{amsmath}
\usepackage{amsthm}
\usepackage{amsfonts}
\usepackage{amssymb}
\usepackage{url}
\usepackage{hyperref}
\usepackage{eucal}
\usepackage{mathrsfs}
\usepackage{graphicx,graphics}
\usepackage[T1]{fontenc}%
\usepackage{url}

\newtheorem{theorem}{Theorem}[section]

\newcommand\fig[1] {{\rm Figure}~\ref{fig:#1}}
\newcommand\labfig[1] {\label{fig:#1}}
\newcommand\sect[1] {\ref{sect:#1}}
\newcommand\labsect[1] {\label{sect:#1}}

\newcommand{\bfm}[1]{\mbox{\boldmath ${#1}$}}
\newcommand{\nonum}{\nonumber \\}
\newcommand\eq[1] {(\ref{#1})}
\newcommand{\beqa}{\begin{eqnarray}}
\newcommand{\eeqa}[1]{\label{#1}\end{eqnarray}}
\newcommand{\beq}{\begin{equation}}
\newcommand{\eeq}[1]{\label{#1}\end{equation}}

\newcommand{\lang}{\langle}
\newcommand{\rang}{\rangle}

\newcommand{\und}[1]{\smash{\underline{#1}}}

\newcommand{\Ga}{\alpha}
\newcommand{\Gb}{\beta}
\newcommand{\Gd}{\delta}
\newcommand{\Ge}{\epsilon}

\newcommand{\Gg}{\gamma}

\newcommand{\Gk}{\kappa}

\newcommand{\Gn}{\eta}
\newcommand{\Gm}{\mu}

\newcommand{\GP}{\Pi}

\newcommand{\BGe}{\bfm\epsilon}

\newcommand{\BGs}{\bfm\sigma}

\newcommand{\CV}{{\cal V}}

\newcommand{\bpm}{\begin{pmatrix}}
\newcommand{\epm}{\end{pmatrix}}

\def\b0{\bf 0}

\def\cof{{\rm cof}}

\def\Ba{{\bf a}}

\def\Bn{{\bf n}}

\def\Bv{{\bf v}}

\def\Bx{{\bf x}}

\def\BA{{\bf A}}
\def\BB{{\bf B}}
\def\BC{{\bf C}}

\def\BI{{\bf I}}

\title{Towards a complete characterization of the effective elasticity tensors of mixtures of an elastic phase and an almost rigid phase}
\date{}
\begin{document}
\maketitle
\vskip -.5cm
\centerline{\large
Graeme W. Milton \footnote{Department of Mathematics, University of Utah, USA -- milton@math.utah.edu,},
\quad
Davit Harutyunyan \footnote{Department of Mathematics, University of Utah, USA -- davith@math.utah.edu,},
\,\, and \,\,
Marc Briane\footnote{Institut de Recherche Math\'ematique de Rennes, INSA de Rennes, FRANCE -- mbriane@insa-rennes.fr,},}
\vskip 1.cm
\begin{abstract}
The set $GU_f$ of possible effective elastic tensors of composites built from two materials with positive definite elasticity tensors $\BC_1$ and $\BC_2=\Gd\BC_0$
comprising the set $U=\{\BC_1,\Gd\BC_0\}$ and mixed in proportions $f$ and $1-f$ is 
partly characterized in the limit $\Gd\to\infty$. The material with tensor $\BC_2$ corresponds to
a material which (for technical reasons) is almost rigid in the limit $\Gd\to \infty$. The paper, and the underlying microgeometries, have many
aspects in common with the companion paper "On the possible effective elasticity tensors of 2-dimensional printed materials". The chief difference is that one has a different algebraic problem to solve: determining the subspaces of
stress fields for which the thin walled structures can be rigid, rather than determining, as in the companion paper, 
the subspaces of strain fields for which the thin walled structure is compliant.
Recalling that $GU_f$ is completely characterized through minimums of sums of energies, involving a set of
applied strains, and complementary energies, involving a set of applied stresses, we provide descriptions of 
microgeometries that in appropriate limits achieve the minimums in many cases. In these cases the calculation of the minimum is reduced to a finite dimensional
minimization problem that can be done numerically. Each microgeometry consists of a union of walls in appropriate directions, where the material
in the wall is an appropriate $p$-mode material, that is almost rigid to $6-p\leq 5$ independent applied stresses, yet is compliant to any strain
in the orthogonal space. Thus the walls, by themselves, can support stress with almost no deformation. The region outside the walls contains ``Avellaneda material'' that
is a hierarchical laminate which minimizes an appropriate sum of elastic energies. 
\end{abstract}
\section{Introduction}
\setcounter{equation}{0}

This paper is a companion to the paper ``On the possible effective elasticity tensors of 2-dimensional and 3-dimensional printed materials'' 
\cite{Milton:2016:PEE}, where a partial characterization
is given  of the set  $GU_f$  of effective elasticity tensors that can be produced in the limit $\Gd\to 0$ if we mix in prescribed proportions $f$ and $1-f$ two materials with
positive definite and bounded elasticity tensors $\BC_1$ and $\BC_2=\Gd\BC_0$. Here we consider the opposite limit $\Gd\to \infty$ which corresponds to mixing in prescribed proportions an elastic phase and an almost rigid phase. Our results are summarized in the theorem in the conclusion section. For a complete introduction and summary of previous results the reader is urged to read at least the first three sections of the companion paper. The essential ideas presented
here are much the same as contained in the companion paper. However the algebraic problem relevant to this paper, of determining when the set of walls can support a set of stress fields, is quite different than the algebraic problem
encountered in the companion paper, of determining when the set of walls is complaint to a set of strain fields.

The microstructures we consider involve taking three limits. First, as they have structure on multiple length scales, the homogenization limit where the ratio between
length scales goes to infinity needs to be taken. Second, the limit $\Gd\to \infty$ needs to be taken. Third, as the structure involves this walls of width $\Ge$, which are very stiff to certain applied stresses, the limit $\Ge\to 0$ needs to be taken so the contribution to the elastic energy of these walls goes to zero, when the structure is compliant to an applied strain. The limits should be taken in this order, as, for example, standard homogenization theory is justified
only if $\Gd$ is positive and finite, so we need to take the homogenization limit before taking the limit $\Gd\to \infty$.

As in the companion paper we emphasize that our analysis is valid only for linear elasticity, and ignores nonlinear effects  such as buckling, which may be important even for small deformations. It is also important to emphasize that 
to our apply our results when phase 2 is perfectly rigid, rather than almost rigid, requires special care. Indeed if phase 2 is perfectly rigid, then many of the microgeometries considered here do not permit the kind of motions that are permitted
for any finite value of $\Gd$, no matter how large. In particular, the structures considered in figures 5, 7, and 8(d) of the
companion paper would be completely rigid if phase 2 was perfectly rigid. To permit the required motions, one has to 
first replace the rigid phase 2 with a composite with a small amount of phase1 as the matrix phase, so that its effective
elastcity tensor is finite, but approaches infinity as the proportion of phase 1 in it tends to zero. The microgeometry in this composite needs to be much smaller than the scales in the geometries discussed here, which would involve mixtures of it and phase 1.

\section{Characterizing $G$ closures through sums of energies and complementary energies}
\setcounter{equation}{0}
Cherkaev and Gibiansky \cite{Cherkaev:1992:ECB, Cherkaev:1993:CEB} found that bounding sums of energies and complementary energies could lead to very useful
bounds on $G$-closures. It was subsequently proved by Francfort and Milton \cite{Francfort:1994:SCE, Milton:1994:LBS} 
that minimums over $\BC_*\in GU_f$ of such sums of energies and complementary energies completely characterize
$GU_f$ in much the same way that Legendre transforms characterize convex sets: the stability under lamination of $GU_f$ is what allows one to recover $GU_f$ from the values
of these minimums (see also Chapter 30 in \cite{Milton:2002:TOC}).
Specifically, in the case of three-dimensional elasticity, the set $GU_f$ is completely characterized if we know the $7$ ``energy functions'',
\beqa W_f^0(\BGs^0_1,\BGs^0_2,\BGs^0_3,\BGs^0_4,\BGs^0_5,\BGs^0_6)
& = &\min_{\BC_*\in GU_f}\sum_{j=1}^6\BGs^0_j:\BC_*^{-1}\BGs^0_j, \nonum
 W_f^1(\BGs^0_1,\BGs^0_2,\BGs^0_3,\BGs^0_4,\BGs^0_5,\BGe^0_1)
& = &\min_{\BC_*\in GU_f}\left[\BGe^0_1:\BC_*\BGe^0_1+\sum_{j=1}^5\BGs^0_j:\BC_*^{-1}\BGs^0_j\right],\nonum
 W_f^2(\BGs^0_1,\BGs^0_2,\BGs^0_3,\BGs^0_4,\BGe^0_1,\BGe^0_2)
& = &\min_{\BC_*\in GU_f}\left[\sum_{i=1}^2\BGe^0_i:\BC_*\BGe^0_i+\sum_{j=1}^4\BGs^0_j:\BC_*^{-1}\BGs^0_j\right],\nonum
 W_f^3(\BGs^0_1,\BGs^0_2,\BGs^0_3,\BGe^0_1,\BGe^0_2,\BGe^0_3)
& = &\min_{\BC_*\in GU_f}\left[\sum_{i=1}^3\BGe^0_i:\BC_*\BGe^0_i+\sum_{j=1}^3\BGs^0_j:\BC_*^{-1}\BGs^0_j\right],\nonum
 W_f^4(\BGs^0_1,\BGs^0_2,\BGe^0_1,\BGe^0_2,\BGe^0_3,\BGe^0_4)
& = &\min_{\BC_*\in GU_f}\left[\sum_{i=1}^4\BGe^0_i:\BC_*\BGe^0_i+\sum_{j=1}^2\BGs^0_j:\BC_*^{-1}\BGs^0_j\right],\nonum
 W_f^5(\BGs^0_1,\BGe^0_1,\BGe^0_2,\BGe^0_3,\BGe^0_4,\BGe^0_5)
& = & \min_{\BC_*\in GU_f}\left[\left(\sum_{i=1}^5\BGe^0_i:\BC_*\BGe^0_i\right)+\BGs^0_1:\BC_*^{-1}\BGs^0_1 \right],\nonum
 W_f^6(\BGe^0_1,\BGe^0_2,\BGe^0_3,\BGe^0_4,\BGe^0_5,\BGe^0_6)
& = & \min_{\BC_*\in GU_f}\sum_{i=1}^6\BGe^0_i:\BC_*\BGe^0_i.
\eeqa{2.1}
In fact, Milton and Cherkaev \cite{Milton:1995:WET} show it suffices
to know these functions for sets of applied strains $\BGe^0_i$ and applied stresses $\BGs^0_j$ that are mutually orthogonal:
\beq (\BGe^0_i,\BGs^0_j)=0,\quad\quad (\BGe^0_i,\BGe^0_k)=0,\quad\quad (\BGs^0_j,\BGs^0_\ell)=0 \quad{\rm for~all~}i, j, k,\ell~~{\rm with}~i\ne j, ~
i\ne k, ~j\ne \ell.
\eeq{2.1aa}
The terms appearing in the minimums have a physical significance. For example, in the expression for $W_f^2$,
\beq \sum_{i=1}^2\BGe^0_i:\BC_*\BGe^0_i+\sum_{j=1}^4\BGs^0_j:\BC_*^{-1}\BGs^0_j \eeq{2.1a}
has the physical interpretation as being the sum of energies per unit volume stored in the composite with effective elasticity tensor $\BC_*$
when it is subjected to successively the two applied strains $\BGe^0_1$ and $\BGe^0_2$ and then to the four applied stresses $\BGs^0_1$, $\BGs^0_2$, $\BGs^0_3$,
$\BGs^0_4$. To distinguish the terms $\BGe^0_i:\BC_*\BGe^0_i$ and $\BGs^0_j:\BC_*^{-1}\BGs^0_j$, the first is called an energy (it is really an energy per unit volume
associated with the applied strain $\BGe^0_i$) and the second is called a complementary energy, although it too physically represents an energy per unit volume
associated with the applied stress $\BGs^0_j$.

For well-ordered materials with $\BC_2\geq\BC_1$ (or the reverse) Avellaneda \cite{Avellaneda:1987:OBM} showed there exist sequentially layered laminates of finite rank
having an effective elasticity tensor $\BC_*=\BC_f^A(\BGe^0_1,\BGe^0_2,\BGe^0_3,\BGe^0_4,\BGe^0_5,\BGe^0_6)$ 
(not to be confused with the elasticity tensor $\BC_*=\widetilde{\BC}_f^A(\BGs^0_1,\BGs^0_2,\BGs^0_3,\BGs^0_4,\BGs^0_5,\BGs^0_6)$ used in the companion paper)
that attains the minimum in the above expression for $W_f^6(\BGe^0_1,\BGe^0_2,\BGe^0_3,\BGe^0_4,\BGe^0_5,\BGe^0_6)$, i.e.,
\beq W_f^6(\BGe^0_1,\BGe^0_2,\BGe^0_3,\BGe^0_4,\BGe^0_5,\BGe^0_6)
 =  \sum_{i=1}^6\BGe^0_i:\BC_f^A(\BGe^0_1,\BGe^0_2,\BGe^0_3,\BGe^0_4,\BGe^0_5,\BGe^0_6)\BGe^0_i.
\eeq{2.1.0.0}
The effective tensor $\BC_*=\BC_f^A(\BGe^0_1,\BGe^0_2,\BGe^0_3,\BGe^0_4,\BGe^0_5,\BGe^0_6)$ of the Avellaneda material is found by finding a combination
of the parameters entering the formula for the effective tensor of sequentially layered laminates that minimizes the sum of six elastic energies. In general this has to be done numerically, but it suffices to consider laminates of rank at most six if $\BC_1$ is isotropic \cite{Francfort:1995:FOM}, or, using an argument of
Avellaneda \cite{Avellaneda:1987:OBM}, to consider laminates of rank at most 21 if $\BC_1$ is anisotropic (see Section 2 in the companion paper).

In the case of two-dimensional elasticity,  the set $GU_f$ is similarly completely characterized if we know the $4$ ``energy functions'',
\beqa
W_f^0(\BGs^0_1,\BGs^0_2,\BGs^0_3,)
& = &\min_{\BC_*\in GU_f}\sum_{j=1}^3\BGs^0_j:\BC_*^{-1}\BGs^0_j, \nonum
 W_f^1(\BGs^0_1,\BGs^0_2,\BGe^0_1)
& = &\min_{\BC_*\in GU_f}\left[\BGe^0_1:\BC_*\BGe^0_1+\sum_{j=1}^2\BGs^0_j:\BC_*^{-1}\BGs^0_j\right],\nonum
 W_f^2(\BGs^0_1,\BGe^0_1,\BGe^0_2)
& = & \min_{\BC_*\in GU_f}\left[\left(\sum_{i=1}^2\BGe^0_i:\BC_*\BGe^0_i\right)+\BGs^0_1:\BC_*^{-1}\BGs^0_1 \right],\nonum
 W_f^3(\BGe^0_1,\BGe^0_2,\BGe^0_3)
 & = & \min_{\BC_*\in GU_f}\sum_{i=1}^3\BGe^0_i:\BC_*\BGe^0_i.
\eeqa{2.40}
Again $W_f^3(\BGe^0_1,\BGe^0_2,\BGe^0_3)$ is attained for an ``Avellaneda material'' consisting of a sequentially layered laminate geometry
having an effective tensor $\BC_*=\BC_f^A(\BGe^0_1,\BGe^0_2,\BGe^0_3)$, i.e.,
\beq W_f^3(\BGe^0_1,\BGe^0_2,\BGe^0_3)=\sum_{i=1}^3\BGe^0_i:\BC_f^A(\BGe^0_1,\BGe^0_2,\BGe^0_3)\BGe^0_i.
\eeq{2.40.0a}
The effective tensor $\BC_*=\BC_f^A(\BGe^0_1,\BGe^0_2,\BGe^0_3)$ of the Avellaneda material is found by finding a combination
of the parameters entering the formula for the effective tensor of sequentially layered laminates that minimizes the sum of three elastic energies. In general this has to be done numerically, but it suffices to consider laminates of rank at most three if $\BC_1$ is isotropic \cite{Avellaneda:1989:BEE}, or, using an argument of
Avellaneda \cite{Avellaneda:1987:OBM}, to consider laminates of rank at most 6 if $\BC_1$ is anisotropic (see Section 2 in the companion paper).

\section{Microgeometries which are associated with sharp bounds on many sums of energies and complementary energies}
\setcounter{equation}{0}
The analysis here of mixtures of an almost rigid phase mixed with an elastic phase is very similar to the analysis in the companion paper for
mixtures of an extremely compliant phase and an elastic phase.  The roles of stresses and strains are interchanged and
now the challenge is to identify matrix pencils that are spanned by matrices with zero determinant, rather than symmetrized rank-one matrices.
We now have the inequalities
\beqa
0 & \leq & W_f^0(\BGs^0_1,\BGs^0_2,\BGs^0_3,\BGs^0_4,\BGs^0_5,\BGs^0_6), \nonum
\BGe^0_1:[\BC_f^A(0, 0, 0, 0, 0, \BGe^0_1)]\BGe^0_1 & \leq & W_f^1(\BGs^0_1,\BGs^0_2,\BGs^0_3,\BGs^0_4,\BGs^0_5,\BGe^0_1), \nonum
\sum_{i=1}^2\BGe^0_i:[\BC_f^A(0, 0, 0, 0,\BGe^0_1,\BGe^0_2)]\BGe^0_i& \leq & W_f^2(\BGs^0_1,\BGs^0_2,\BGs^0_3,\BGs^0_4,\BGe^0_1,\BGe^0_2), \nonum
\sum_{i=1}^3\BGe^0_i:[\BC_f^A(0, 0, 0, \BGe^0_1,\BGe^0_2,\BGe^0_3)]\BGe^0_i& \leq & W_f^3(\BGs^0_1,\BGs^0_2,\BGs^0_3,\BGe^0_1,\BGe^0_2,\BGe^0_3), \nonum
\sum_{i=1}^4\BGe^0_i:[\BC_f^A(0, 0, \BGe^0_1,\BGe^0_2,\BGe^0_3,\BGe^0_4)]\BGe^0_i &\leq & W_f^4(\BGs^0_1,\BGs^0_2,\BGe^0_1,\BGe^0_2,\BGe^0_3,\BGe^0_4), \nonum
\sum_{i=1}^5\BGe^0_i:[\BC_f^A(0,\BGe^0_1,\BGe^0_2,\BGe^0_3,\BGe^0_4,\BGe^0_5)]\BGe^0_i & \leq & W_f^5(\BGs^0_1,\BGe^0_1,\BGe^0_2,\BGe^0_3,\BGe^0_4,\BGe^0_5). 
\eeqa{5.1}
The first inequality is clearly sharp, being attained when the composite consists of islands of phase 1 surrounded by a phase 2 (so that $(\BC_*)^{-1}$ approaches $0$ as $\Gd\to\infty$).
Again the objective is to show that many of the other inequalities are sharp too in the limit $\Gd\to\infty$ at least when the spaces spanned by the applied stresses
$\BGs^0_j$, $j=1,2,\ldots,6-p$ satisfy certain properties. This space of applied stresses associated with $W_f^p$ has dimension $6-p$ and its orthogonal complement defines
the $p$ dimensional space $\CV_p$.

The recipe for doing this is to simply insert into a relevant Avellaneda material a microstructure occupying a thin walled region containing a $p$--mode material, such that the wall structure, by itself, is very stiff when the applied stress lies in the $(6-p)$-dimensional subspace spanned by the $\BGs^0_j$, yet allows
strains in the orthogonal $p$-dimensional subspace $\CV_p$ spanned by the $\BGe^0_i$. We say a composite with effective tensor $\BC_*$ built from the two materials $\BC_1$ and $\BC_2=\Gd\BC_0$ is very stiff to a stress $\BGs^0_j$ if the complementary energy $\BGs^0_j:\BC_*^{-1}\BGs^0_j$ goes to zero as
$\Gd\to \infty$, and allows a strain $\BGe^0_i$ if the elastic energy $\BGe^0_i:\BC_*\BGe^0_i$ has a finite limit as $\Gd\to \infty$. These $p$--mode materials
have exactly the same construction as that specified in Section 5.3 of the companion paper, only now the region that was occupied by the elastic phase is now occupied
by the rigid phase, and the material that was occupied by the extremely compliant phase (which becomes void in the limit $\Gd\to 0$) is occupied by the elastic phase.
If we happened to choose $\BC_0=\BC_1$ all the moduli (and effective moduli) are simply rescaled, i.e., for any $\Gd$, and in particular for large values of $\Gd$, if a mixture of two materials with effective tensors $\BC_1$ and $\BC_1/\Gd$ has effective tensor $\BC_*$, then when rescale the elasticity tensors of the
two phases to
$\Gd\BC_1$ and $\BC_1$, the resulting effective elasticity tensor will be $\Gd\BC_*$. Thus the analysis of the response of the $p$--mode materials is essentially the same as in the companion paper. Exactly the same trial fields can be chosen to bound the response of the  $p$--mode material. Hence we
will not repeat this analysis but instead the reader is referred to Section 5.3 of the companion paper.

The subspace orthogonal to $\CV_p$ is now required to be 
spanned by matrices $\BGs^{(k)}$, $k=1,\ldots,6-p$ such that
\beq \BGs^{(k)}\Bn_k=0, \eeq{5.1a}
for some unit vector $\Bn_k$. Thus the identifying feature of these matrices  $\BGs^{(k)}$ is that they have zero determinant, and then $\Bn_k$ can be chosen as a null-vector of  $\BGs^{(k)}$.
The existence of such matrices $\BGs^{(k)}$ is proved in Section~\sect{det0symmat}. The proof uses small perturbations of the applied stresses and strains.
But, due the continuity of the energy functions $W_f^k$ established in Section~\sect{continuity}, the small perturbations do not modify the generic result.
The vectors $\Bn_k$ determine the orientation of the walls in the structure since a set of walls orthogonal to $\Bn$ can support
any stress $\BGs$ such that $\BGs\Bn=0$. 

To define the thin walled structure, introduce the periodic function $H_c(x)$ with period $1$ which takes the value $1$ 
if $x-[x]\leq c$ where $[x]$ is the greatest integer less than $x$, and $c\in[0,1]$ gives the relative thickness of each wall.
 Then for the unit vectors
$\Bn_1,\Bn_2,\ldots,\Bn_{6-p}$ appearing in \eq{5.1a}, and for a small relative thickness $c=\Ge$ define the characteristic functions 
\beq \Gn_k(\Bx)=H_\Ge(\Bx\cdot\Bn_k+k/p). \eeq{2.13}
This characteristic function defines a series of parallel walls, as shown in \fig{2}(a),
each perpendicular to the vector $\Bn_j$, where $\Gn_j(\Bx)=1$ in the wall material.  The additional shift term $k/p$ in \eq{2.13} ensures the walls
associated with $k_1$ and $k_2$ do not intersect when it happens that $\Bn_{k_1}=\Bn_{k_2}$, at least when $\Ge$ is small. We emphasize that $\Ge$ is not a homogenization parameter, but rather represents a volume fraction of walls.

Now define the characteristic 
function
\beq \chi_*(\Bx)=\prod_{k=1}^{p}(1-\Gn_k(\Bx)). \eeq{2.14}
If $p\leq 3$ this is usually a periodic function of $\Bx$ -- an exception being if $p=3$ and there are no nonzero integers $z_1$, $z_2$, and $z_3$
such that $z_1\Bn_1+z_2\Bn_2+z_3\Bn_3=0$. More generally, $\chi_*(\Bx)$ is a quasiperiodic function of $\Bx$. 
The walled structure is where $\chi_*(\Bx)$ takes the value $0$. In the case $p=2$ the wall structure is illustrated in \fig{2}(b).

\begin{figure}[!ht]
\centering
\includegraphics[width=0.7\textwidth]{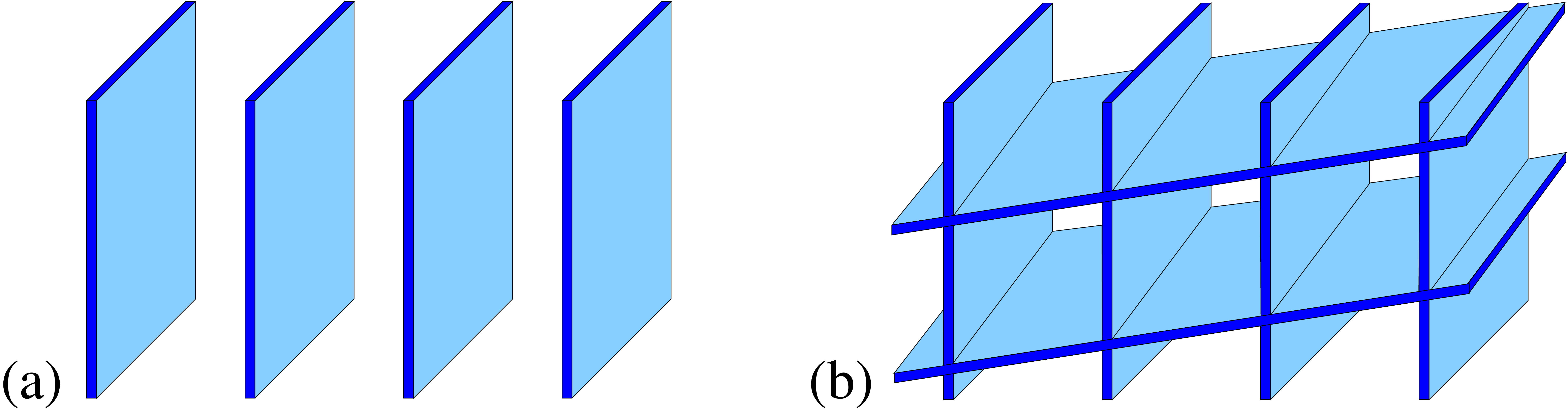}
\caption{Example of walled structures. In (a) we have a ``rank 1'' walled structure and in (b) a ``rank 2'' walled structure. The generalization to walled
structures of any rank is obvious, and precisely defined by the characteristic function \eq{2.14} that is 0 in the walls, and 1 in the remaining material.}
\labfig{2}
\end{figure}

The walled structure is where $\chi_*(\Bx)$ given by \eq{2.14} takes the value $0$.
 Inside it we put a $p$--mode material with effective tensor $\BC_*^{2}=\BC_*(\CV_p)$
that allows any applied strain $\BGe^0$ in the space $\CV_p$ but which is very stiff
to any stress $\BGs^0$ orthogonal to the space $\CV_p$. Using the 6 matrices
\beq  \Bv_1=\BGs_1^0/|\BGs_1^0|,\ldots,\Bv_{6-p}=\BGs_{6-p}^0/|\BGs_{6-p}^0|, \Bv_{7-p}=\BGe^0_1/|\BGe^0_1|,\ldots,\Bv_6=\BGe^0_p/|\BGe^0_p|,
\eeq{2.15} 
as our basis for the $6$-dimensional space of $3\times 3$ symmetric matrices
the compliance tensor $[\BC_*(\CV_p)]^{-1}$ in this basis takes the limiting form
\beq \lim_{\Gd\to\infty}[\BC_*(\CV_p)]^{-1}=\bpm
0 & 0 \\ 0 & \BB
\epm
\eeq{5.2}
where $\BB$ represents a (strictly) positive definite $p\times p$ matrix and the $0$ on the diagonal represents the $(6-p)\times (6-p)$ zero matrix. 
Inside the walled structure, where $\chi_*(\Bx)=1$ we put the Avellaneda material with effective elasticity tensor
\[
\BC_*^{1}=\BC_f^A(0,\ldots,0,\BGe^0_1,\ldots,\BGe^0_p).
\]

In a variational principle similar to (4.4) in the companion paper (i.e., treating the Avellaneda material and the $p$--mode material both as homogeneous 
materials with effective tensors $\BC_*^{1}=\BC_f^A$ and $\BC_*^{2}=\BC_*(\CV_p)$, respectively) we choose trial strain fields that are constant:
\beq \und{\BGe}_i(\Bx)=\BGe^0_i,\quad{\rm for}~i=1,2,\ldots,p, \eeq{5.3}
thus trivially fulfilling the differential constraints, and trial stress fields of the form
\beq \und{\BGs}_j(\Bx)=\sum_{k=1}^{6-p}\BGs_{j,k}\Gn_k(\Bx)/\Ge, \eeq{5.4}
which are required to have the average values 
\beq \BGs^0_j=\lang \und{\BGs}_j\rang=\sum_{k=1}^{6-p}\BGs_{j,k}, \eeq{5.5}
and the matrices $\BGs_{i,j}$ are additionally required to lie in the space orthogonal to $\CV_p$ (so they cost very little energy) and satisfy
\beq \BGs_{j,k}=c_{j,k}\BGs^{(k)}, \eeq{5.6}
for some choice of parameters $c_{j,k}$ to ensure that $\BGs_{j,k}\Bn_k=0$ and hence that
$\und{\BGs}_j(\Bx)$ satisfies the differential constraints of a stress field-- this requires $\und{\BGs}_j(\Bx)\Bn_k$ to be continuous across
any interface with normal $\Bn_k$. Additionally, the $c_{j,k}$ in \eq{5.6} should be chosen so the $\BGs^0_j$ given by
\eq{5.5} are orthogonal.

To find upper bounds on the energy associated with this trial stress field, first consider those parts of the wall structure that are outside of any junction regions, i.e., where for some $k$, $\Gn_k(\Bx)=1$, while $\Gn_s(\Bx)=0$ for all $s\ne k$. An upper bound for the volume fraction occupied by the region where $\Gn_k(\Bx)=1$ while $\Gn_s(\Bx)=0$ for all $s\ne k$ is of course $\Ge$ as this represents the 
volume of the region where $\Gn_k(\Bx)=1$. The associated energy per unit volume of the trial stress field in those parts of the wall structure that are outside of any junction regions is bounded above by
\beq \sum_{k=1}^{6-p} \BGs_{j,k}:[\BC_*(\CV_p)]^{-1}\BGs_{j,k}/\Ge. \eeq{2.22a}
With an appropriate choice of multimode material one can construct bounded trial stress fields that are 
essentially concentrated in phase 2 and consequently $\BGs_{j,k}:[\BC_*(\CV_p)]^{-1}\BGs_{j,k}$ is bounded 
above by a quantity proportional to $1/\Gd$. Our assumption that we take the limit $\Gd\to\infty$ before taking
the limit $\Ge\to 0$ ensures that  $1/(\Gd\Ge)\to 0$, and thus ensures that the quantity \eq{2.22a} goes to zero in this limit.

Next, consider those junction regions where only two walls meet, i.e., where for some $k_1$ and $k_2> k_1$, $\Bx$  is such that $\Gn_{k_1}(\Bx)=\Gn_{k_2}(\Bx)=1$ 
while $\Gn_s(\Bx)=0$ for all  $s$ not equal to $k_1$ or $k_2$. Provided $\Bn_{k_1}\ne\Bn_{k_2}$, an upper bound for the volume fraction occupied by each such junction region is $\Ge^2$. Then the associated energy per unit volume of the trial stress field in these junction regions where only two walls meet is bounded above by
\beq \sum_{k_1=1}^{6-p} \sum_{k_2=k_1+1}^{6-p} (\BGs_{i,k_1}+\BGs_{i,k_2}):[\BC_*(\CV_p)]^{-1}(\BGs_{j,k_1}+\BGs_{j,k_2}). \eeq{2.22b}
Thus the powers of $\Ge$ cancel and this energy density will go to zero if the multimode material is easily compliant to the strains $\BGs_{j,k_1}+\BGs_{j,k_2}$ for all $k_1$ and $k_2$ with $k_2>k_1$. 

Finally, consider those junction regions where three or more walls meet, i.e., for some $k_1$, $k_2> k_1$, and $k_3> k_2$, $\Bx$  is such that $\Gn_{k_i}(\Bx)=1$ for $i=1,2,3$. For a given
choice of $k_1$, $k_2> k_1$, and $k_3> k_2$ such that the
three vectors  $\Bn_{k_1}$, $\Bn_{k_2}$, and $\Bn_{k_3}$ are not coplanar an upper bound for the volume fraction occupied by this region is $\Ge^3$. In the case that the three vectors 
$\Bn_{k_1}$, $\Bn_{k_2}$, and $\Bn_{k_3}$ are coplanar, we can ensure that the  volume fraction occupied by this region is  $\Ge^3$ or less by appropriately translating one or two wall structures, i.e., by replacing $\Gn_{k_m}(\Bx)$ with $\Gn_{k_m}(\Bx+\Ga_i\Bn_{k_m})$ for $m=2,3$, for an appropriate choice of $\Ga_2$ and $\Ga_3$ between $0$ and $1$. Since the energy density of the trial field
in these regions scales as $\Ge^3/\Ge^2=\Ge$ we can ignore this contribution in the limit $\Ge\to 0$ as it goes to zero too.

From this analysis of the energy densities associated with the trial fields it follows that one does not necessarily need the pentamode, quadramode, trimode, bimode, and unimode materials as appropriate for the material inside the walled structure. Instead, by modifying the construction, it suffices to use only pentamode and quadramode materials. In the walled structure we now put pentamode materials in those sections where for some $k$, $\Gn_{k}(\Bx)=1$ while $\Gn_{k'}(\Bx)=0$ for 
all $k'\ne k$. Each pentamode material is very stiff to the single stress $\BGs^{(k)}$ appropriate to the wall under consideration.
 In each junction region of the walled structure where $\Gn_{k_1}(\Bx)=\Gn_{k_2}(\Bx)=1$ for some $k_1\ne k_2$ 
while $\Gn_{k}(\Bx)=0$ for all $k$ not equal to $k_1$ or $k_2$, we put a quadramode material which is very stiff to any stress in the subspace spanned by 
$\BGs^{(k_1)}$ and $\BGs^{(k_2)}$ as appropriate to the 
junction region under consideration.
 In the remaining junction regions of the walled
structure (where three or more walls intersect) we put phase 1. The contribution to the average energy of the fields in these regions 
vanishes as $\Ge\to 0$ as discussed above.  

By these constructions we effectively obtain materials with elasticity tensors $\BC_*$ such that
\beq \lim_{\Gd\to \infty}(\BC_*)^{-1}=\GP_p(\BC_f^A)^{-1}\GP_p, \eeq{5.6a}
where $\BI$ is the fourth-order identity matrix, $\GP_p$ is the fourth-order tensor that is the projection onto the space $\CV_p$, and $\BC_f^A$ is the relevant Avellaneda material.
In the basis \eq{2.15} $\GP_p$ is represented by the 6$\times$6 matrix that has the block form, 
\beq \GP_p=\bpm
0 & 0 \\ 0 & \BI_p
\epm,
\eeq{5.6b}
where $\BI_p$ represents the $p\times p$ identity matrix and the $0$ on the diagonal represents the $(6-p)\times (6-p)$ zero matrix.

In the case $d=2$ the analysis simplifies in the obvious way. We have the inequalities
\beqa
0 & \leq & W_f^0(\BGs^0_1,\BGs^0_2,\BGs^0_3), \nonum
\BGe^0_1:[\BC_f^A(0, 0, \BGe^0_1)]\BGe^0_1 & \leq & W_f^1(\BGs^0_1,\BGs^0_2,\BGe^0_1), \nonum
\sum_{i=1}^2\BGe^0_i:\BC_f^A(0, \BGe^0_1,\BGe^0_2)\BGe^0_i &\leq & W_f^2(\BGs^0_1,\BGe^0_1,\BGe^0_2),
\eeqa{5.7}
the first one of which is sharp in the limit $\Gd\to\infty$ being attained when the material consists of islands of phase 1 surrounded by phase 2.
The recipe for showing that the bound \eq{5.7} on $W_f^1(\BGs^0_1,\BGs^0_2,\BGe^0_1)$ is sharp for certain values of $\BGs^0_1$ and $\BGs^0_2$ and that the bound \eq{2.40} on 
$W_f^2(\BGs^0_1,\BGe^0_1,\BGe^0_2)$ is sharp for certain values of $\BGs^0_1$ is almost exactly the same as in the $3$-dimensional case:
insert into the Avellaneda material a thin walled structure of respectively unimode and bimode materials so that it is very stiff to any 
stress in the space spanned by $\BGs^0_1$ and $\BGs^0_2$ in the case of $W_f^1$, or so that it is very stiff to the stress $\BGs^0_1$ in the case of $W_f^2$.

\section{The algebraic problem: characterizing those symmetric matrix pencils spanned by zero determinant matrices}
\labsect{det0symmat}
\setcounter{equation}{0}

Now we are interested in the following question: \textit{Given $k$ linearly independent symmetric $d\times d$ matrices $\BA_1,\BA_2,\dots,\BA_k$, find necessary and sufficient conditions such, that there exists linearly independent matrices $\{\BB_i\}_{i=1}^k$ spanned by the basis elements $\BA_i$ such that $\det(\BB_i)=0.$
It is assumed that $d=2$ or $3$ and $1\leq k\leq k_d,$ where $k_2=2$ and $k_3=5.$}
Here, we are working in the generic situation, i.e., we prove the algebraic result for a dense set of matrices. The continuity result of Section 5 will allow us to conclude for the whole set of matrices. Actually, the proof below also shows that the algebraic result holds for the complementary of a zero measure set of matrices. Let us prove the following theorem.

\begin{theorem}
The above problem is solvable if and only if the matrices $\BA_i,$ $i=1,\dots,k$ satisfy the following condition:
\begin{itemize}
\item[(i)]
\beq
\det(\BA_1)=0,\quad\text{if}\quad k=1,\  d=2,3.
\eeq{6.1}
\item[(ii)]
\beq
(\Ga_1\Gg_2+\Ga_2\Gg_1-2\Gb_1\Gb_2)^2>4\det(\BA_1)\det(\BA_2),\quad\text{if}\quad k=d=2,
\eeq{6.2}
where
\beq \BA_i=
\bpm
\Ga_i & \Gb_i\\
\Gb_i & \Gg_i
\epm
\eeq{6.3}
\item[(iii)]
\beqa
\triangle=18\det(\BA_1)\det(\BA_2)S_1S_2-4S_1^3\det(\BA_2)+S_1^2S_2^2-4S_2^3\det(\BA_1)\nonum
\kern -2.cm -27\det(\BA_1)^2\det(\BA_2)^2>0\qquad\qquad\qquad\qquad \text{if}\quad k=2,\ d=3,
\eeqa{6.4}
where $S_i=\sum_{j=1}^3{s_{ij}},$ $i=1,2$ and $s_{ij}$ is the determinant of the matrix obtained by replacing the $j-$th row of $\BA_i$ by the $j-$th row of $\BA_{i+1},$ where by convention we have $\BA_3=\BA_1.$

\item[(iv)]
\beq
\text{Always solvable if}\quad k\geq 3,\ d=3.
\eeq{6.5}
\end{itemize}
\end{theorem}
\noindent {\bf Remark.}
In fact the condition \eq{6.1} that $\det(\BA_1)=0$ could be excluded since we are considering the generic case. It is inserted because we can treat it explicitly.
\begin{proof}
We consider all the cases separately. \\
\textbf{Case (i): $k=1.$} In this case one must obviously have $\det(\BA_1)=0$.\\
\textbf{Case (ii): $k=d=2.$} We can without loss of generality assume, that (by small perturbations) $\det(\BA_i)\neq 0, i=1,2.$
For $\Gn,\Gm\in\mathbb R^2$, denote $\BA(\Gn,\Gm)=\Gn \BA_1+\Gm \BA_2,$ and thus for the equality
\beq \det(\BA(\Gn,\Gm))=\det(\BA_1)\Gn^2+(\Ga_1\Gg_2+\Ga_2\Gg_1-2\Gb_1\Gb_2)\Gn\Gm+\det(\BA_2) \Gm^2 \eeq{6.6}
to happen, one must first of all have $ \Gm\neq 0,$ thus dividing by $ \Gm^2$ and denoting $t={\Gn}/{\Gm},$ we get that the
quadratic equation
\beq \frac{1}{ \Gm^2}\det(\BA(\Gn,\Gm))=\det(\BA_1)t^2+(\Ga_1\Gg_2+\Ga_2\Gg_1-2\Gb_1\Gb_2)t+\det(\BA_2)=0, \eeq{6.7}
must have two different solutions, i.e., the discriminant is strictly positive, which amounts to exactly \eq{6.2}.\\
\textbf{Case (iii): $k=2,\ d=3.$} Again, we can without loss of generality assume, that $\det(\BA_i)\neq 0, i=1,2.$ Denote then again
$\BA(\Gn,\mu)=\Gn \BA_1+\mu \BA_2,$ thus we must have, that the equation
\beq
\det(\BA(\Gn,\mu))=\det(\BA_1)\Gn^3+S_1\Gn^2\mu+S_2\Gn\mu^2+\det(\BA_2)\mu^3=0
\eeq{6.8}
has at least two different real roots, which gives by Cardan's condition
\beq
\begin{array}{ll}
\triangle = & 18\det(\BA_1)\det(\BA_2)S_1S_2-4S_1^3\det(\BA_2)+S_1^2S_2^2-4S_2^3\det(\BA_1)
\\*[.2em]
& -\,27\det(\BA_1)^2\det(\BA_2)^2>0,
\end{array}
\eeq{6.9}
which is exactly \eq{6.4}.\\
\textbf{Case (iv): $k\geq 3,\ d=3.$} Let us consider the case $k=3$ first. Let us show, that we can assume without loss of generality, that
$\det(\BA_1)=\det(\BA_2)=0,$ by proving, that there exist numbers $\Gn_i\neq 0,\ i=1,2$ such, that
the matrices $\BB_1=\Gn_1 \BA_1+\BA_2,$ $\BB_2=\Gn_2 \BA_1+\BA_3$ have zero determinant.
Indeed, we assume without loss of generality, that $\det(\BA_i)\neq 0,\ i=1,2,3.$ We would like to have then
\beq \det(\BB_1)(\Gn_1)=\Gn_1^3\det(\BA_1)+\Gn_1^2(\cdot)+\Gn_1(\cdot)+\det(\BA_2)=0, \eeq{6.10}
which has a nonzero root $\Gn_1$ being a cubic equation and as $\det(\BB_1)(0)=\det(\BA_2)\neq 0.$ Similarly, the equation
 $\det(\BB_2)(\Gn_2)=0$ has a nonzero solutions $\Gn_2.$ The matrices $\BB_1,\BB_2$ and $\BA_1$ are linearly independent, because the linear independence of $\BB_1,\BB_2$ and $\BA_1$ is equivalent to the condition
\beq \det
 \bpm
 \Gn_1 & 1 & 0\\
 \Gn_2 & 0 & 1\\
 1 & 0 & 0
  \epm=1\neq 0. \eeq{6.11}
Assume now that $\BA_1$, $\BA_2$ and $\BA_3$ are linearly independent and
\beq
\det(\BA_1)=\det(\BA_2)=0.
\eeq{6.12}
For any $\Gn,\mu\in\mathbb R,$ consider the matrix $\BB_3=\BB(\Gn,\mu)=\BA_3+\Gn \BA_1+\mu \BA_2.$ It is clear, that the triple $\BA_1,\BA_2,\BB_3$
is linearly independent, so we would like to show that there exist $\Gn,\mu\in\mathbb R,$ such that $\det(\BB_3)=0.$ Assume in contradiction, that
\beq
\det(\BB_3)\neq 0,\quad\text{for all}\quad \Gn,\mu\in\mathbb R.
\eeq{6.13}
Let us then show, that the condition \eq{6.13} implies that $c_1=c_2=0,$ where taking into account the condition \eq{6.12} we have that
\beq
\det(\BB_3)=c_1\Gn^2\mu+c_2\Gn\mu^2+c_3\Gn\mu+c_4\Gn^2+c_5\mu^2+c_6\Gn+c_7\mu+\det(\BA_3).
\eeq{6.14}
Indeed, if $c_1\neq 0$ then taking $\Gn=\mu^2$ we get that the equation $\det(\BB(\mu^2,\mu))=0$ would have a solution $\mu\in\mathbb R,$ being a fifth order equation, thus we get $c_1=c_2=0.$ Next, by perturbing the elements of $\BA_1$ and $\BA_2$ if necessary, we can reach the situation where no entries and second order minors of both $\BA_1$ and $\BA_2$ vanish, by first reaching the situation when $\BA_1$ and $\BA_2$ have no zero entries. If we now perturb any $ij$ and $ik$ elements of $\BA_1$ by small numbers $\epsilon$ and $\delta,$ where $j\neq k,$ then to keep the condition
$\det(\BA_1)=0,$ so $\epsilon$ and $\delta$ must satisfy the relation 
\beq
\epsilon\cdot \cof_{ij}(\BA_1)+\delta\cdot\cof_{ik}(\BA_1)=0.
\eeq{6.15}
On the other hand, the condition $c_2=0$ must not be violated by that perturbation, thus we must have as well
\beq
\epsilon\cdot \cof_{ij}(\BA_2)+\delta\cdot\cof_{ik}(\BA_2)=0.
\eeq{6.16}
The last two conditions then imply that the cofactor matrix $\cof{\BA_1}$ is a multiple of the cofactor matrix $\cof{\BA_2},$ i.e.,
\beq
\cof(\BA_2)=t\cdot\cof(\BA_1),\quad t\neq 0.
\eeq{6.17}
Again, a small perturbation of the $11$ and $12$ elements of $\BA_1$ by $\epsilon$ and $\delta$ satisfying \eq{6.15} with $i=j=1, k=2$ does not violate the condition $\det(\BA_1)=0$, thus it must not violate the condition \eq{6.16}. Observe, that the above perturbation does not change the cofactor
$\cof_{11}(\BA_1),$ but it changes the cofactor element $\cof_{33}(\BA_1),$ which means, that the desired condition $\det(\BB_3)=0$ can be reached by small perturbations. The case $k=d=3$ is now done.\\
Assume now $k\geq 4$ and $d=3.$ By the previous step, in the space spanned by $\BA_1$, $\BA_2$, and $\BA_3$ there are three matrices $\BA_1'$, $\BA_2'$ and $\BB_3=\BA_3+\Gn_3\BA_1'+\mu_3\BA_2'$ that are linearly independent matrices with zero determinant. Then again by the previous step we can find linearly independent matrices $\BB_1,\ldots,\BB_k$ that have the form 
$\BB_1=\BA_1',$ $\BB_2=\BA_2'$ and $\BB_i=\BA_i+\Gn_i\BA_1'+\mu_i\BA_2'$ for $3\leq i\leq k$, that are linearly independent and have zero determinant.
Thus the proof is finished.
\end{proof}
\section{Continuity of the energy functions}
\labsect{continuity}
\setcounter{equation}{0}
It follows from the preceding analysis that we can determine the three energy functions
$W_f^1(\BGs^0_1,\BGs^0_2,\BGs^0_3,\BGs^0_4,\BGs^0_5,\BGe^0_1)$,
$W_f^2(\BGs^0_1,\BGs^0_2,\BGs^0_3,\BGs^0_4,\BGe^0_1,\BGe^0_2)$, and
$W_f^3(\BGs^0_1,\BGs^0_2,\BGs^0_3,\BGe^0_1,\BGe^0_2,\BGe^0_3)$
in the limit $\Gd\to \infty$ for almost all combinations of applied fields.
Here we establish that these energy functions are continuous functions of the
applied fields in the limit
$\Gd\to \infty$, and therefore we obtain expressions for the energy functions for all combinations of applied fields in this limit.

Recall that the set $G_fU$ is characterized by its $W$-transform. For example, part of it is described by the function
\beq W_f^2(\BGs^0_1,\BGs^0_2,\BGs^0_3,\BGs^0_4,\BGe^0_1,\BGe^0_2)
 = \min_{\BC_*\in GU_f}\left[\sum_{i=1}^2\BGe^0_i:\BC_*\BGe^0_i+\sum_{j=1}^4\BGs^0_j:\BC_*^{-1}\BGs^0_j\right],
\eeq{8.1}
Here we want to show that such energy functions are continuous in their arguments.
Let the compliance tensor $[\BC_*(\BGs^0_1,\BGs^0_2,\BGs^0_3,\BGs^0_4,\BGe^0_1,\BGe^0_2)]^{-1}$ 
be a minimizer of \eq{8.1}, and suppose we perturb the applied stress fields $\BGs^0_j$ by $\Gd\BGs^0_j$,
and the applied strain fields $\BGe^0_i$ by $\Gd\BGe^0_i$. Now consider the following
walled material, with a geometry described by the characteristic function
\beq \chi_w(\Bx)=\prod_{k=1}^{3}(1-H_{\Ge'}(\Bx\cdot\Bn_k)), \eeq{8.2}
where $\Bn_1$, $\Bn_2$, and $\Bn_3$ are the three orthogonal unit vectors,
\beq \Bn_1=\bpm 1 \\ 0 \\ 0 \epm,\quad \Bn_2=\bpm 0 \\ 1 \\ 0 \epm,\quad\Bn_3=\bpm 0 \\ 0 \\ 1 \epm, \eeq{8.3}
and $\Ge'$ is a small parameter that gives the thickness of the walls. Inside the walls, where $\chi_w(\Bx)=0$ we put an
isotropic composite of phase 1 and phase 2, mixed in the proportions $f$ and $1-f$ with isotropic effective
elasticity tensor $\BC(\Gk_0,\Gm_0)$, where $\Gk_0$ is the effective bulk modulus and $\Gm_0$ is the effective 
shear modulus, that are assumed to have finite limits as $\Gd\to \infty$. (The isotropic composite could consist of islands of void
surrounded by phase 1). Outside the walls, where $\chi_w(\Bx)=1$, we put the material that has effective compliance tensor 
$[\BC_*^{1}]^{-1}=[\BC_*(\BGs^0_1,\BGs^0_2,\BGs^0_3,\BGs^0_4,\BGe^0_1,\BGe^0_2)]^{-1}$. Let $\BC_*'$ be the effective tensor of the composite. We have the variational
principle 
\beqa &~&\sum_{i=1}^2(\BGe^0_i+\Gd\BGe^0_i):\BC_*'(\BGe^0_i+\Gd\BGe^0_i)+\sum_{j=1}^4(\BGs^0_j+\Gd\BGs^0_j):(\BC_*')^{-1}(\BGs^0_j+\Gd\BGs^0_j)=\nonum
&~&\min_{\und{\BGe}_1,\und{\BGe}_2,\und{\BGe}_3,\und{\BGe}_4,\und{\BGs}_1,\und{\BGs}_2}
\Big\langle\sum_{i=1}^2\und{\BGe}_i(\Bx):[\chi_w(\Bx)\BC_*^{1}+(1-\chi_w(\Bx))\BC(\Gk_0,\Gm_0)]\und{\BGe}_i(\Bx)\nonum
&~&+\sum_{j=1}^4\und{\BGs}_j(\Bx):[\chi_w(\Bx)\BC_*^{1}+(1-\chi_w(\Bx))\BC(\Gk_0,\Gm_0)]^{-1}\und{\BGs}_j(\Bx)\Big\rangle,
\eeqa{8.4}
where the minimum is over fields subject to the appropriate average values and differential constraints. We choose 
constant trial stress fields
\beq \und{\BGs}_j(\Bx)=\BGs^0_j+\Gd\BGs^0_j,\quad j=1,2,3,4, \eeq{8.5}
and trial strain fields
\beq \und{\BGe}_i(\Bx)=\BGe^0_i+\Gd\und{\BGe}_i(\Bx),\quad i=1,2, \eeq{8.6}
where $\Gd\und{\BGe}_i(\Bx)$ has average value $\Gd\BGe^0_i$ and is concentrated in the walls. Specifically, if $\{\Gd\BGe^0_i\}_{k\ell}$
denote the matrix elements of $\Gd\BGe^0_i$, and letting
\beqa \Gd\BGe_i^1& = &\bpm \{\Gd\BGe^0_i\}_{11} & \{\Gd\BGe^0_i\}_{12} & 0 \\ \{\Gd\BGe^0_i\}_{21} & 0 & 0 \\ 0 & 0 & 0 \epm,\nonum
\Gd\BGe_j^2 & = &\bpm 0 & 0 & 0 \\ 0 & \{\Gd\BGe^0_i\}_{22} & \{\Gd\BGe^0_i\}_{23} \\ 0 & \{\Gd\BGe^0_i\}_{32} & 0 \epm, \nonum
\Gd\BGe_j^3 & = &\bpm 0 & 0 &\{\Gd\BGe^0_i\}_{13} \\ 0 & 0 & 0 \\ \{\Gd\BGe^0_i\}_{31} & 0 & \{\Gd\BGe^0_i\}_{33} \epm, 
\eeqa{8.7}
then we choose
\beq \Gd\und{\BGe}_i(\Bx)=\sum_{k=1}^{3}\Gd\BGe_i^k H_{\Ge'}(\Bx\cdot\Bn_k)/\Ge', \eeq{8.8}
which has the required average value $\Gd\BGs^0_j$ and satisfies the differential constraints appropriate to a strain field because
$\Gd\BGe_i^k=\Ba_{i,k}\Bn_k^T+\Bn_k\Ba_{i,k}^T$ for some vector $\Ba_{i,k}$.

Hence there
exist constants $\Ga$ and $\Gb$ such that for sufficiently small $\Ge'$ and for sufficiently small variations $\Gd\BGs^0_j$ and $\Gd\BGe^0_i$ in the applied fields,
we have
\beqa &~&\Big\langle\sum_{i=1}^2\und{\BGe}_i(\Bx):[\chi_w(\Bx)\BC_*^{1}+(1-\chi_w(\Bx))\BC(\Gk_0,\Gm_0)]\und{\BGe}_i(\Bx)\nonum
&~&+\sum_{j=1}^4\und{\BGs}_j(\Bx):[\chi_w(\Bx)\BC_*^{1}+(1-\chi_w(\Bx))\BC(\Gk_0,\Gm_0)]^{-1}\und{\BGs}_j(\Bx)\Big\rangle \nonum
&~&\leq W_f^2(\BGs^0_1,\BGs^0_2,\BGs^0_3,\BGs^0_4,\BGe^0_1,\BGe^0_2)+\Ga\Ge'+\Gb K/\Ge'
\eeqa{8.8a}
where $K$ represents the norm 
\beq K=\sqrt{\sum_{i=1}^2\Gd\BGe^0_i:\Gd\BGe^0_i+\sum_{j=1}^4\Gd\BGs^0_j:\Gd\BGs^0_j}, \eeq{8.9}
of the field variations. Choosing $\Ge'=\sqrt{\Gb K/\Ga}$ to minimize the right hand side of \eq{8.8a} we obtain
\beqa &~& W_f^2(\BGs^0_1+\Gd\BGs^0_1,\BGs^0_2+\Gd\BGs^0_2,\BGs^0_3+\Gd\BGs^0_3,\BGs^0_4+\Gd\BGs^0_4,\BGe^0_1+\Gd\BGe^0_1,\BGe^0_2+\Gd\BGe^0_2)\nonum
&~&\quad\quad\leq W_f^2(\BGs^0_1,\BGs^0_2,\BGs^0_3,\BGs^0_4,\BGe^0_1,\BGe^0_2)+2\sqrt{\Ga\Gb K}.
\eeqa{8.10}
Clearly the right hand side approaches $W_f^2(\BGs^0_1,\BGs^0_2,\BGs^0_3,\BGs^0_4,\BGe^0_1,\BGe^0_2)$ as $K\to 0$. On the other hand
by repeating the same argument with the roles of $W_f^2(\BGs^0_1,\BGs^0_2,\BGs^0_3,\BGs^0_4,\BGe^0_1,\BGe^0_2)$ and
$W_f^2(\BGs^0_1+\Gd\BGs^0_1,\BGs^0_2+\Gd\BGs^0_2,\BGs^0_3+\Gd\BGs^0_3,\BGs^0_4+\Gd\BGs^0_4,\BGe^0_1+\Gd\BGe^0_1,\BGe^0_2+\Gd\BGe^0_2)$ reversed,
and with the compliance tensor $[\BC_*(\BGs^0_1+\Gd\BGs^0_1,\BGs^0_2+\Gd\BGs^0_2,\BGe^0_1+\Gd\BGe^0_1,\BGe^0_2+\Gd\BGe^0_2,\BGe^0_3+\Gd\BGe^0_3,\BGe^0_4+\Gd\BGe^0_4)]^{-1}$
replacing the compliance tensor $[\BC_*(\BGs^0_1,\BGs^0_2,\BGs^0_3,\BGs^0_4,\BGe^0_1,\BGe^0_2)]^{-1}$ we deduce that
\beqa &~& W_f^2(\BGs^0_1,\BGs^0_2,\BGs^0_3,\BGs^0_4,\BGe^0_1,\BGe^0_2) \nonum
&~&\quad\leq W_f^2(\BGs^0_1+\Gd\BGs^0_1,\BGs^0_2+\Gd\BGs^0_2,\BGs^0_3+\Gd\BGs^0_3,\BGs^0_4+\Gd\BGs^0_4,\BGe^0_1+\Gd\BGe^0_1,\BGe^0_2+\Gd\BGe^0_2)
+2\sqrt{\Ga\Gb K}.\nonum &~&
\eeqa{8.11}
This with \eq{8.10} establishes the continuity of $W_f^2(\BGs^0_1,\BGs^0_2,\BGs^0_3,\BGs^0_4,\BGe^0_1,\BGe^0_2)$. The continuity of the other energy functions follows by
the same argument. 
\subsection{Conclusion}
\setcounter{equation}{0}
To conclude, we have established the following Theorems:
\begin{theorem}
Consider composites in three dimensions of two materials with positive definite elasticity tensors $\BC_1$ and $\BC_2=\Gd\BC_0$ mixed in proportions $f$ and $1-f$. Let the seven energy functions $W_f^k$, $k=0,1,\ldots,6$, that characterize the set $G_fU$ (with $U=(\BC_1,\Gd\BC_0)$) of possible elastic tensors
be defined by \eq{2.1}. These  energy functions involve a set of applied strains $\BGe^0_i$ and applied stresses $\BGs^0_j$ meeting the orthogonality condition \eq{2.1aa}. The energy function $W_f^6$ is given by
\beq
 W_f^6(\BGe^0_1,\BGe^0_2,\BGe^0_3,\BGe^0_4,\BGe^0_5,\BGe^0_6)
 = \sum_{i=1}^6\BGe^0_i:\BC_f^A(\BGe^0_1,\BGe^0_2,\BGe^0_3,\BGe^0_4,\BGe^0_5,\BGe^0_6)\BGe^0_i,
\eeq{7.8}
as established by Avellaneda \cite{Avellaneda:1987:OBM}, where $\BC_f^A(\BGe^0_1,\BGe^0_2,\BGe^0_3,\BGe^0_4,\BGe^0_5,\BGe^0_6)$ is the effective elasticity tensor of an Avellaneda material, that is a sequentially layered laminate with the minimum value of the sum of elastic energies 
\beq \sum_{i=1}^6\BGe^0_j:\BC_*\BGe^0_j.
\eeq{7.8a}
Again some of the applied stresses $\BGs_j^0$ or applied strains $\BGe^0_i$ could be zero. Additionally we have
\beqa
 \lim_{\Gd\to \infty}W_f^0(\BGs^0_1,\BGs^0_2,\BGs^0_3,\BGs^0_4,\BGs^0_5,\BGs^0_6) & = & 0,\nonum
 \lim_{\Gd\to \infty}W_f^1(\BGs^0_1,\BGs^0_2,\BGs^0_3,\BGs^0_4,\BGs^0_5,\BGe^0_1) & = & \BGe^0_1:[\BC_f^A(0, 0, 0, 0, 0, \BGe^0_1)]\BGe^0_1, \nonum
\lim_{\Gd\to \infty}W_f^2(\BGs^0_1,\BGs^0_2,\BGs^0_3,\BGs^0_4,\BGe^0_1,\BGe^0_2)   & = & \sum_{i=1}^2\BGe^0_i:[\BC_f^A(0, 0, 0, 0,\BGe^0_1,\BGe^0_2)]\BGe^0_i, \nonum
\lim_{\Gd\to \infty} W_f^3(\BGs^0_1,\BGs^0_2,\BGs^0_3,\BGe^0_1,\BGe^0_2,\BGe^0_3)   &= &\sum_{i=1}^3\BGe^0_i:[\BC_f^A(0, 0, 0, \BGe^0_1,\BGe^0_2,\BGe^0_3)]\BGe^0_i,
\eeqa{7.9}
for all combinations of applied stresses $\BGs_j^0$ and applied strains $\BGe^0_i$. When $\det(\BGs^0_1)=0$ we have
\beq
\lim_{\Gd\to \infty} W_f^5(\BGs^0_1,\BGe^0_1,\BGe^0_2,\BGe^0_3,\BGe^0_4,\BGe^0_5)=\sum_{i=1}^5\BGe^0_i:[\BC_f^A(0,\BGe^0_1,\BGe^0_2,\BGe^0_3,\BGe^0_4,\BGe^0_5)]\BGe^0_i,
\eeq{7.10}
while, when $f(t)=\det(\BGs^0_1+t\BGs^0_2)$ has at least two roots (the condition for which is given by \eq{6.4}), 
\beq
\lim_{\Gd\to \infty}W_f^4(\BGs^0_1,\BGs^0_2,\BGe^0_1,\BGe^0_2,\BGe^0_3,\BGe^0_4)=\sum_{i=1}^4\BGe^0_i:[\BC_f^A(0, 0, \BGe^0_1,\BGe^0_2,\BGe^0_3,\BGe^0_4)]\BGe^0_i.
\eeq{7.11}
\end{theorem}
\begin{theorem}
For two-dimensional composites the four energy functions $W_f^k$, $k=0,1,2,3$ are defined by \eq{2.40} and these characterize the set $G_fU$, with $U=(\BC_1,\Gd\BC_0)$, of possible elastic tensors $\BC_*$ of composites of two phases with positive definite elasticity tensors $\BC_1$ and
$\BC_2=\Gd\BC_0$.  The energy functions involve a set of applied strains $\BGe^0_i$ and applied stresses $\BGs^0_j$ meeting the orthogonality condition \eq{2.1aa}. The energy function $W_f^3$ is given by
\beq
 W_f^3(\BGe^0_1,\BGe^0_2,\BGe^0_3)
 = \sum_{i=1}^3\BGe^0_i:\BC_f^A(\BGe^0_1,\BGe^0_2,\BGe^0_3)\BGe^0_i,
\eeq{7.12}
as proved by Avellaneda \cite{Avellaneda:1987:OBM}, where $\BC_f^A(\BGe^0_1,\BGe^0_2,\BGe^0_3)$ is the effective elasticity tensor of an Avellaneda material, that is a sequentially layered laminate with the minimum value of the sum of elastic energies 
\beq \sum_{j=1}^3\BGe^0_j:\BC_*\BGe^0_j.
\eeq{7.12a}
We also have the trivial result that
\beq
 \lim_{\Gd\to \infty}W_f^0(\BGs^0_1,\BGs^0_2,\BGs^0_3)=0.
\eeq{7.13}
When $\det\BGs^0_1=0$, we have
\beq
 \lim_{\Gd\to \infty}W_f^2(\BGs^0_1,\BGe^0_1,\BGe^0_2)=\sum_{i=1}^2\BGe^0_i:[\BC_f^A(0,\BGe^0_1,\BGe^0_2)]\BGe^0_i,
\eeq{7.14}
while when $f(t)=\det(\BGs^0_1+t\BGs^0_2)$ has exactly two roots (the condition for which is given by \eq{6.2}),
\beq
\lim_{\Gd\to \infty}W_f^1(\BGs^0_1,\BGs^0_2,\BGe^0_1)=\BGe^0_1:[\BC_f^A(0, 0, \BGe^0_1)]\BGe^0_1.
\eeq{7.15}
\end{theorem}

These theorems, and the accompanying microstructures, help define what sort of elastic behaviors are theoretically possible in 2-d and 3-d materials consisting
of a very stiff phase and an elastic phase (possibly anisotropic, but with fixed orientation).
They should serve as benchmarks for the construction of more realistic microstructures that can be manufactured. We have found the minimum over all microstructures of various sums of energies and complementary energies. 

It remains an open problem to find expressions for the energy functions in the cases not covered by these theorems.
Notice that for three-dimensional composites the function $W_f^5$ is only determined when special condition $\det(\BGs^0_1)=0$ is satisfied exactly. Similarly, for two-dimensional composites the function $W_f^2$ is only
determined when the special condition $\det\BGs^0_1=0$ is satisfied exactly. Thus these functions are only known on a set of zero measure.

 Even for an isotropic composite with
a bulk modulus $\Gk_*$ and a shear modulus $\Gm_*$, the set of all possible pairs $(\Gk_*, \Gm_*)$ is still not completely characterized either in the limit
$\Gd\to\infty$. In these limits the bounds of Berryman and Milton \cite{Berryman:1988:MRC} 
and Cherkaev and Gibiansky \cite{Cherkaev:1993:CEB} decouple and provide no extra information 
beyond that provided by the Hashin-Shtrikman-Hill bounds \cite{Hashin:1963:VAT, Hashin:1965:EBF, Hill:1963:EPR, Hill:1964:TMP}.  While the results of this paper show that in the limit $\Gd\to \infty$  one can obtain 
three-dimensional structures attaining the Hashin-Shtrikman-Hill lower bound on $\Gk_*$, while having $\Gm_*=\infty$, it is not clear what the minimum value for $\Gm_*$ is, given that
$\Gk_*=\infty$, nor is it clear in two-dimensions what is the minimum value of $\Gk_*$, when $\Gm_*=\infty$.

\section*{Acknowledgements}
The authors thank the National Science Foundation for support through grant DMS-1211359. M. Briane wishes to thank the Department of Mathematics of the University of Utah for his stay during March 25-April 3 2016.
\bibliographystyle{siamplain}
\bibliography{/u/ma/milton/tcbook,/u/ma/milton/newref}
\end{document}